\documentstyle[preprint,aps]{revtex}
\begin{document}
\renewcommand{\theequation}{\arabic{section}.\arabic{equation}}
\draft
\title{
\rightline{\rm \normalsize JHU-TIPAC-940008}
\rightline{\rm \normalsize INFNCA-TH-94-8}
\vspace*{2.0cm}
GENERATIONAL MASS SPLITTING OF NEUTRINOS IN HIGH TEMPERATURE
$SU(2)_{\scriptscriptstyle{L}}\otimes U(1)$ GAUGE THEORY}
\author{A. Erdas}
\address{
Dipartimento di Scienze Fisiche, Universit\`a di Cagliari,
09124 Cagliari, Italy\\
I.N.F.N. Sezione di Cagliari, 09127 Cagliari, Italy}
\author{C. W. Kim and J. A. Lee}
\address{
Department of Physics and Astronomy \\
The Johns Hopkins University, Baltimore, MD 21218
\vspace*{1.5cm}}

\maketitle
\begin {abstract} We calculate the generational mass splitting of
neutrinos in high temperature $SU(2)_{\scriptscriptstyle{L}}\otimes U(1)$ gauge
theory when the temperature is above $250$ GeV
and the gauge symmetry is restored.
We consider the case of neutrinos that are massless at tree level as well as
the case of neutrinos with tree-level mass and large mixing.
\vspace*{-0.3truecm}

\noindent \underline{$\phantom{space space space space}$}

\noindent
 erdas@cagliari.infn.it; cwkim@jhuvms.hcf.jhu.edu; leej@dirac.pha.jhu.edu
\end {abstract}
\newpage
\setcounter{equation}{0}
\section{Introduction}

As neutrinos play a significant role in the early universe
\cite{kimpevsner93},
their behavior in hot and dense media has been extensively investigated
by many authors.
Recently, several of them extended their interest
into the $SU(2)_{\scriptscriptstyle{L}}\otimes U(1)$ symmetric
phase of the universe where
the gauge particles associated with the symmetry are massless while
the fermions need not be massless.

We have examined the contributions of the charged leptonic background to
the neutrino self-energy, from which we calculate the neutrino effective mass,
in the era where the $SU(2)_{\scriptscriptstyle{L}}\otimes U(1)$ symmetry
is unbroken. Since the $SU(2)_{\scriptscriptstyle{L}}\otimes U(1)$ symmetry
of the Lagrangian prohibits the presence of a mass term for the fermions,
here we are
referring to the running masses which are
obtained by running the low energy fermion masses to high energy,
using the renormalization group equation.
The temperature is assumed to be not much higher than
250 GeV so only the Standard Model leptons are relevant.

The calculation shows that, unlike what happens in the vacuum case,
the neutrinos, which are massless
at the tree level, acquire effective masses that depend on the generation
indices.
In other words, the degeneracy among the massless neutrinos is removed
through the interactions with the charged leptons.
This can be of immediate interest since even a slight generational
asymmetry among
the neutrinos can lead to resonant neutrino oscillations \cite{Samuel}
and change the evolution of the early universe.

We show in Section II the detailed calculation of the neutrino self-energy
\begin{equation}\Sigma(K)=-a\gamma_\mu K^{\mu} -b\gamma_\mu u^{\mu}\, ,
\label{E:1_1}\end{equation}
where $u^\alpha$ is the four-velocity of the heat bath,
$K^\alpha$ is the four-momentum of the neutrino under consideration,
and $a$ and $b$
are Lorentz-invariant functions that depend on the Lorentz-scalars
 defined by\cite{Weldon}
\begin{equation}\omega\equiv K^\alpha u_\alpha,\;\;\;\;\; k\equiv
\left[(K^\alpha u_\alpha
)^2-K^2\right]^{1/2}\,.
\label{E:1_2}\end{equation}
In Section III we then find the poles of the neutrino propagator
\begin{equation}S(K)=[\gamma_\mu K^{\mu}+\Sigma(K)]^{-1}=
[(1+a)\gamma_\mu K^{\mu}+b\gamma_\mu u^{\mu}]^{-1}\,.
\label{E:1_3}\end{equation}
and obtain the mass differences between two different species of neutrinos.
We also discuss the case when the neutrinos are mixed and have non vanishing
tree level masses at zero temperature.
Finally we conclude the article giving a summary of the work and its
implications on the early universe.

\setcounter{equation}{0}
\section{ Computation of the self-energy}

We are considering the $SU(2)_{\scriptscriptstyle{L}}
\otimes U(1)$ gauge theory when the
temperature is above $250$ GeV and the gauge symmetry is restored.
We are interested in calculating the finite temperature corrections
to the neutrino
self-energy at the one-loop level using the real-time
formulation of finite-temperature field theory. We will neglect
the effects of a possibly non-vanishing (but still very small) $CP$-asymmetry
in the fermionic background. The three kinds of Feynman
diagrams that are relevant to this calculation are shown in Fig. 1.
We take the
neutrinos to be massless Dirac particles and the charged
leptons to have ``running masses''.

The issue of the fermion masses must be carefully addressed at this point.
In this work we are referring to the fermion running masses. The fermion
running masses in the energy region where GUTS are broken but
$SU(2)_{\scriptscriptstyle{L}}\otimes U(1)$ is still an exact
symmetry can be obtained by
using the renormalization group equations to run the values of the fermion
masses from their values at the low energy $\mu$ to the values at the higher
energy $p$. The running fermion
masses are
\begin{equation}m_\alpha(p^2)=m_\alpha(\mu^2)\left[{1\over 1+{g^2(\mu^2)\over
4\pi}
{4\over 3\pi}\ln\left({p^2\over\mu^2}\right)}\right]^{(3/4)^3}
\left[{1\over 1-{g'^2(\mu^2)\over 4\pi}
{1\over 2\pi}\ln\left({p^2\over\mu^2}\right)}\right]^{-(9/20)}\,\,,
\label{E:2_1}\end{equation}
where $m_\alpha(\mu^2)$ is the charged lepton mass as it is measured in
the lab at low energy (i.e. $\mu$ = 1 GeV), $g$ is the
$SU(2)_{\scriptscriptstyle{L}}$ coupling constant, $g'$ is the $U(1)$
coupling constant, and $\alpha =e,\mu,\tau$.
We also need to clarify that these running masses do not include
any matter effects, since it is well known that the fermion effective
mass due to matter effect is of order $gT$, where $T$ is the temperature
\cite{Weldon}.
Now we can write the finite temperature propagators for fermions $S(p)$
and gauge
bosons $D_{\mu \nu}(p)$ in the Feynman gauge as
\begin{equation}S(p)=[\gamma_\mu p^{\mu}+m(p^2)]\left[{1\over p^2-
m^2(p^2)+i\epsilon}+
i\Gamma_f (p)\right]\,\,,
\label{E:2_2}\end{equation}
\begin{equation}D_{\mu \nu}(p)=-\eta_{\mu \nu}\left[{1\over p^2+i\epsilon}-
i\Gamma_b (p)\right]\,\,,
\label{E:2_3}\end{equation}
where $m(p^2)$ is the fermion mass of Eq. (\ref{E:2_1}), and
$\Gamma_f (p)$ and $\Gamma_b (p)$ are defined as
\begin{equation}\Gamma_f (p)=2\pi \delta (p^2-m^2(p^2)) n_f(p)
\label{E:2_5}\end{equation}
\begin{equation}\Gamma_b (p)=2\pi \delta (p^2) n_b(p)
\label{E:2_4}\end{equation}
with
\begin{equation}n_f(p)=\left[ e^{|p \cdot u |/T} +1 \right]^{-1}
\label{E:2_7}\end{equation}
\begin{equation}n_b(p)=\left[ e^{|p \cdot u |/T} -1 \right]^{-1}\ .
\label{E:2_6}\end{equation}

The imaginary part of the finite temperature correction
to the neutrino self-energy can be neglected \cite{Weldon,us}, and therefore
the contribution from
the tadpole diagrams will vanish. Since the calculations of the gauge boson
contribution and of the scalar contribution are very similar, we will show
only some details of the first one.

The gauge boson diagrams that
contribute to the one loop neutrino self-energy are the $W$-lepton
diagram, $\Sigma_{(1)}$, and the $Z$-neutrino diagram, $\Sigma_{(2)}$,
and they are given by
\begin{equation}\Sigma_{(n)}(K)=ig^2C_{(n)} \int {d^4p\over(2\pi)^4}
D_{\mu \nu}(p) \gamma^{\mu}S_{(n)}(p+K)
\gamma^{\nu}\ ,
\label{E:2_8}\end{equation}
where the index $n$ refers to 1 or 2, $g$ is the coupling
constant, the numerical factor $C_{(n)}$ is $C_{(1)}=1/2$
(for the $W$-lepton diagram) and $C_{(2)}=1/4$
(for the $Z$-neutrino diagram),
$S_{(1)}(p+K)$ is the propagator for massive leptons of Eq. (\ref{E:2_2}) and
$S_{(2)}(p+K)$ is the propagator for massless neutrinos which is obtained
by setting $m(p^2)=0$ in Eq. (\ref{E:2_2}).
Once we calculate $\Sigma_{(1)}(K)$, the value
of $\Sigma_{(2)}(K)$ can be obtained from $\Sigma_{(1)}(K)$ by setting
$m(p^2)=0$ and multiplying by 1/2.

The finite temperature
correction $\Sigma'$ to the neutrino self-energy is
	\begin{equation}
	\Sigma'= \Sigma-\Sigma\,(T=0)
	\label{E:2_9}
	\end{equation}
and using the expressions
for the propagators
(\ref{E:2_2}) and (\ref{E:2_3}), we obtain
\begin{equation}
	\Sigma'_{(1)}=2g^2C_{(1)}\int {d^4p\over(2\pi)^4} [\gamma_\mu p^{\mu} +
	\gamma_\mu K^{\mu} -2m(T^2)] \left[
	{\Gamma_b (p) \over (p+K)^2-m^2(T^2)}
	- {\Gamma_f (p+K)\over p^2} \right]
\label{E:2_10}\end{equation}
where we neglected the imaginary part of $\Sigma^\prime$.
In Eq. (\ref{E:2_10}) we replaced $m(p^2)$ with $m(T^2)$,
because the fermion masses depend very slowly
on the fermion four momentum.
The term proportional to $2m$ vanishes because it is sandwiched between
left-handed and right-handed projectors. Changing $p$ to $-p-K$ we
find
\begin{equation}
	\Sigma'_{(1)}=2g^2C_{(1)}\int {d^4p\over(2\pi)^4}
	\left[ {(\gamma_\mu p^{\mu} + \gamma_\mu K^{\mu})
	\Gamma_b (p)\over (p+K)^2-m^2}  +{\gamma_\mu p^{\mu}
	\Gamma_f (p)\over (p+K)^2}  \right]
\label{E:2_11}\end{equation}
where, for brevity, we used $m$ for $m(T^2)$.
After integration over $p_0$ and over the angles, we obtain
\begin{eqnarray}
	{1\over 4}Tr&&\hskip -0.5cm (\gamma_\mu K^{\mu}\Sigma_{(1)}')
\nonumber \\
	&=&g^2C_{(1)}\int
	{dp\over8\pi^2} \biggl\{ \biggl[ 4p- {K^2+m^2\over
	2k}
\nonumber\\
	&&\times\biggl(\ln \left[{\omega^2-k^2-m^2+2(\omega-k)p\over
	\omega^2-k^2-m^2+2(\omega+k)p}\right]
	+ \ln \left[{\omega^2-k^2-m^2+2(-\omega-k)p\over \omega^2-k^2-m^2+2(-
	\omega+k)p}\right] \biggr)\biggl]
	\left[ e^{p/T} -1 \right]^{-1}
\nonumber \\
	&&+\biggl[4p+ {K^2+m^2\over 2k}\biggl(
	\ln \left[{\omega^2-k^2+m^2+2\omega \sqrt{p^2+m^2}-
	2k p\over \omega^2-k^2+m^2+2\omega \sqrt{p^2+m^2}+2k
 	p}\right]
\nonumber \\
	&&+\ln \left[{\omega^2-k^2+m^2-2\omega \sqrt{p^2+m^2}-
	2k p\over \omega^2-k^2+m^2-2\omega \sqrt{p^2+m^2}+2k p}
	\right]\biggr)\biggr]{p\over \sqrt{p^2+m^2}}
 	\left[ e^{(\sqrt{p^2+m^2})/T} +1 \right]^{-1}\biggr\}\,\,,
	\label{E:2_12}\end{eqnarray}

\begin{eqnarray}
	{1\over 4}Tr&&\hskip -0.5cm(\gamma_\mu u^{\mu}\Sigma_{(1)}')
\nonumber \\
	&=&{g^2C_{(1)}\over k}
	\int {dp\over8\pi^2}\biggl\{-(p+\omega)
 	\ln \left[{\omega^2-k^2-m^2+2(\omega-k)p\over
	\omega^2-k^2-m^2+2(\omega+k)p}\right]
	\left[ e^{p/T} -1 \right]^{-1}
\nonumber \\
	&&\phantom{aaaa}-p
	\,\left[ e^{(\sqrt{p^2+m^2})/T} +1
	\right]^{-1}\ln \left[{\omega^2-k^2+m^2+2\omega
	\sqrt{p^2+m^2}-2k p\over
	\omega^2-k^2+m^2+2\omega \sqrt{p^2+m^2}+2k p}\right]
  	\biggr\}
\nonumber \\
	&&+{g^2C_{(1)}\over k}\int {dp\over8\pi^2}\biggl\{ (p-\omega)
 	\ln \left[{\omega^2-k^2-m^2+2(-\omega-k)p\over
	\omega^2-k^2-m^2+2(-\omega+k)p}\right]\left[ e^{p/T} -1 \right]^{-1}
\nonumber \\
	&&\phantom{aaaa}+p \,\ln \left[{\omega^2-k^2+m^2-2\omega
	\sqrt{p^2+m^2}-2k p\over
	\omega^2-k^2+m^2-2\omega \sqrt{p^2+m^2}+2k p}
	\right]\left[ e^{(\sqrt{p^2+m^2})/T} +1 \right]^{-1}  \biggr\}\,\,,
\label{E:2_13}\end{eqnarray}
where all logarithms are to be understood in the principal-value
sense.
\setcounter{equation}{0}
\section{ Calculation of the generational mass-splitting of neutrinos}

The finite temperature correction to the
neutrino self-energy can be written in terms of the two Lorentz-invariant
functions $a$ and $b$ as
	\begin{equation}
	\Sigma^\prime(K)=\Sigma^\prime_{(1)}(K)+\Sigma^\prime_{(2)}(K)=
	-a\gamma_\mu K^{\mu} -b\gamma_\mu u^{\mu}\,\,.
	\label{E:3_1}
	\end{equation}
Therefore we have
\begin{eqnarray}
	{1\over 4}Tr(\gamma_\mu K^{\mu}\Sigma')&=&-aK^2-bK\cdot u=
	-a(\omega^2-k^2)-b\omega
\nonumber \\
	{1\over 4}Tr(\gamma_\mu u^{\mu}\Sigma')&=&-aK\cdot
	 u-b=-a\omega-b\,\,,
	\label{E:3_3}
\end{eqnarray}
from which we obtain
\begin{eqnarray}
	ak^2&=&{1\over 4}Tr(\gamma_\mu K^{\mu}\Sigma')-
	\omega{1\over 4}Tr(\gamma_\mu u^{\mu}\Sigma')
\nonumber \\
	bk^2&=&-\omega {1\over 4}Tr(\gamma_\mu K^{\mu}\Sigma')+
	(\omega^2-k^2){1\over 4}Tr(\gamma_\mu u^{\mu}\Sigma')\,\,.
\label{E:3_5}\end{eqnarray}

The poles in the neutrino propagator occur when $\omega$ and $k$ are such
as to produce a zero in the Lorentz-invariant function $D$
	\begin{equation}D=(1+a)^2K^2+b^2+2(1+a)bK\cdot u\; .
	\label{E:3_6}\end{equation}
One obtains the positive-energy root of $D=0$ when $\omega$
and $k$ satisfy
\begin{equation}\omega(1+a)+b=k(1+a)
\label{E:3_7}\end{equation}
We want to find the solution of this equation for $k=0$, and the
value of  $\omega$ (which we call $M$) that satisfies Eq.
(\ref{E:3_7}), when $k=0$,
is the neutrino effective mass.

For small $k$ the two expressions ${1\over 4}Tr(\gamma_\mu K^{\mu}\Sigma')$
and ${1\over 4}Tr(\gamma_\mu u^{\mu}\Sigma')$ are of the form
\begin{eqnarray}
	{1\over 4}Tr(\gamma_\mu K^{\mu}\Sigma')&=&h_0+h_1k^2+h_2k^4+\cdots
	\nonumber \\
	{1\over 4}Tr(\gamma_\mu u^{\mu}\Sigma')
	&=&g_0+g_1k^2+g_2k^4+\cdots\,\,,
\label{E:3_9}\end{eqnarray}
where the $h_i$ and $g_i$ are functions of $\omega$ and $T$ only,
i.e. $h_i=h_i(\omega,T)$ for $i=0,1,\cdots$ and
$g_i=g_i(\omega,T)$ for $i=0,1,\cdots$

We have shown in our previous paper \cite{us} that the effective neutrino
mass $M$ is given by
\begin{equation}M^2=h_0(\omega=M,T)\,\,.
\label{E:3_11}\end{equation}
The Eq. (\ref{E:3_11}) is valid only if $h_0$ and $g_0$
(as defined in Eq. (\ref{E:3_9})) satisfy the following
	\begin{equation}h_0=\omega g_0\,\, .
	\label{E:3_10}
	\end{equation}
We have checked explicitly that in our case Eq. (\ref{E:3_10}) is satisfied
and therefore we can use Eq. (\ref{E:3_11}) to calculate $M$.

We are interested in temperatures of $250$ Gev or higher, and so we
can take $m\ll T$ in our calculations.
To obtain the value of $h_0$ we use the following integrals
	\begin{equation}\int_0^\infty p\left[ e^{p/T} -1
 	\right]^{-1}dp={\pi^2\over6}T^2
	\label{E:3_12}
	\end{equation}
	\begin{equation}\int_0^\infty {p^2\over\sqrt{p^2+m^2}}\left[
	e^{(\sqrt{p^2+m^2})/T} +1 \right]^{-1}dp=
	{\pi^2\over12}T^2+{1\over 8}m^2
	\left[\ln\left(m^2\over T^2\pi^2\right)+2\gamma_{\scriptscriptstyle{E}}
	-1\right]
	\label{E:3_13}
	\end{equation}
where $\gamma_{\scriptscriptstyle{E}}=0.5772$ and Eq. (\ref{E:3_13})
is valid only if $m\ll T$ \cite{Dolan}.
We then obtain, for $m/T\ll 1$,
\begin{eqnarray}
	\lim_{k\rightarrow 0}&&\hskip -0.5cm{1\over 4}Tr(\gamma_\mu
K^{\mu}\Sigma_{(1)}')
\nonumber \\
&=
	&\phantom{+}
	g^2C_{(1)}{T^2\over 8}+g^2C_{(1)}{1\over 16\pi^2}m^2
		\left[
			\ln(m^2/T^2\pi^2)+
			2\gamma_{\scriptscriptstyle{E}} -1
		\right]
\nonumber \\
&	&+g^2C_{(1)}(\omega^2-m^4/\omega^2)
	\int {dp\over8\pi^2}
		\biggl(-{1\over\omega-m^2/\omega+ 2p}
			+{1\over \omega-m^2/\omega-2p}
		\biggr)
		\left[ e^{p/T} -1 \right]^{-1}
\nonumber\\
	&&	+g^2C_{(1)}(\omega^2-m^4/\omega^2)
	\int_m^\infty {dp\over8\pi^2}
		\biggl({1\over \omega+m^2/\omega+ 2p}
			-{1\over \omega+m^2/\omega- 2p}
		\biggr)
		\left[ e^{p/T} +1 \right]^{-1}\; ,
\nonumber \\
&&\label{E:3_14}
\end{eqnarray}
which, by neglecting terms of order $m^4$ and higher, can be re-written as
\begin{eqnarray}
	\lim_{k\rightarrow 0}{1\over 4}Tr&&\hskip -0.5cm
	(\gamma_\mu K^{\mu}\Sigma_{(1)}')
\nonumber \\
	&=&g^2C_{(1)}{T^2\over 8}+g^2C_{(1)}
	{1\over 16\pi^2}m^2\left[\ln(m^2/T^2
	\pi^2)+2\gamma_{\scriptscriptstyle{E}} -1\right]
\label{E:3_15}\\
	&&+g^2C_{(1)}\omega^2\int_0^\infty {dp\over8\pi^2}
		\biggl(-{1\over\omega+p}+{1\over \omega-p}
		\biggr)
		\left[ e^{p/T} -1 \right]^{-1}
\nonumber \\
	&&+g^2C_{(1)}{m^2\over8\pi^2}
\nonumber \\
	&&+g^2C_{(1)}\omega m^2
	\int_0^\infty {dp\over8\pi^2}
		\biggl[
			{1\over (\omega- 2p)^2}-{1\over (\omega+ 2p)^2}
		\biggr]
		\left\{
			\left[ e^{p/T} +1 \right]^{-1}
			+\left[ e^{p/T} -1 \right]^{-1}
		\right\}
	\,\, .\nonumber
\end{eqnarray}
Similarly, we have
\begin{equation}\lim_{k\rightarrow 0}
{1\over 4}Tr(\gamma_\mu K^{\mu}\Sigma_{(2)}')=
g^2C_{(2)}{T^2\over 8}+g^2C_{(2)}\omega^2\int_0^\infty {dp\over8\pi^2}
\biggl(-{1\over\omega+p}+{1\over \omega-
p} \biggr)
\left[ e^{p/T} -1 \right]^{-1}
\label{E:3_16}\end{equation}
and therefore adding Eqs. (\ref{E:3_15}) and (\ref{E:3_16})
and setting $\omega=M$
we obtain the value of $h_0$, which gives us the following
equation for the neutrino effective mass squared
\begin{eqnarray}
	M^2=M_0^2\biggl\{&&{3\over4}+{1\over 4T^2\pi^2}m^2
	\left[\ln(m^2/T^2\pi^2)+2\gamma_{\scriptscriptstyle{E}} +1\right]
\nonumber \\
	&&+{3\over 4}{M^2\over T^2}\int_0^\infty {dp\over\pi^2}
	\biggl(-{1\over M+p}+{1\over M-
	p} \biggr)
	\left[ e^{p/T} -1 \right]^{-1}
\nonumber \\
	&&+{1\over 2}{M\over T}{m^2\over T^2}\int_0^\infty {dp\over\pi^2}
	\biggl[{1\over (M/T-
	2p)^2}-{1\over (M/T+ 2p)^2}\biggr]
	\left[e^{p}\over e^{2p} -1 \right]\biggr\}\,\,,
\label{E:3_17}\end{eqnarray}
where we have used $M_0^2=g^2{T^2\over 8}$.

When we also add the scalar
contributions to the fermion self-energy,
$M_0^2$ is modified to
	\begin{equation}
	M_0^2={g^2\over 8}T^2+{|f|^2\over 8}T^2\ ,
	\label{E:3_18}
	\end{equation}
where $|f|$ is the Yukawa coupling constant of the fermions
\cite{Weldon,us}. We find that the contribution from the
Yukawa couplings can be neglected for $|f|^2\ll g^2$.

The term on the third line of Eq.(\ref{E:3_17})
can be neglected compared with
the other terms because it is of order $g(m/T)^2$.
The term on the second line of Eq. (\ref{E:3_17})
will contribute
to the effective mass a negligible amount compared to ${3\over4}M_0^2$ (but
still
much more than the $m^2$), but since it is independent of the lepton mass
it will not have any effect on the generational mass splitting.
Therefore we conclude that the
difference between the square of effective mass of
neutrino of the $\alpha$ generation and the square of effective
mass of neutrino of the $\beta$ generation
($\alpha, \beta =e,\mu,\tau$) is given by
\begin{eqnarray}
	\Delta_{\alpha\beta}=M^2_\alpha-M^2_\beta=
	&&\phantom{-}{g^2\over
	32\pi^2}m^2_\alpha(T^2)\Bigl\{
	\ln\left[m^2_\alpha(T^2)/T^2\pi^2\right]+
	2\gamma_{\scriptscriptstyle{E}}
	+1\Bigr\}
\nonumber \\
	&&-{g^2\over 32\pi^2}m^2_\beta(T^2)\Bigl\{
	\ln\left[m^2_\beta(T^2)/T^2\pi^2\right]+
	2\gamma_{\scriptscriptstyle{E}}
	+1\Bigr\}
\label{E:3_20}\end{eqnarray}
where $m_\alpha(T^2)$ and $m_\beta(T^2)$ are the running masses of
the corresponding charged leptons given by Eq. (\ref{E:2_1}).

Now we discuss the case of massive neutrinos
with non-vanishing mixing angles.
In this case the distinction
between the mass and the weak eigenstates is critical
\cite{carlokimlee92}.
The mass eigenstates are the one particle eigenstates of the neutrino
field operators that have mutually independent quadratic (free) Lagrangians.
The weak eigenstates, on the other hand, are the linear
combinations of the mass eigenstates that vanish when
acted upon by the weak fields that are responsible
for the weak interaction.
The weak eigenstates (fields) and the mass eigenstates
(fields) are related by the unitary mixing matrix $U$ as,
\begin{equation}
\mid \nu_i\  > =U_{i\alpha}^\ast \mid \nu_\alpha\ > \ ,\
(\nu_i  =U_{i\alpha}\nu_\alpha)\ ,
\label{E:3_21}\end{equation}
where $\mid \nu_i\ >,\; i=1,2,3$ are the mass eigenstates and
$\mid \nu_\alpha\ >,\; \alpha=e,\mu,\tau$ are the weak eigenstates.
(For more comprehensive discussion, see Ref.\cite{carlokimlee92}.)
By definition, the propagator and the dispersion relation can be defined only
with the mass eigenstates.
So our discussion should be understood in terms of the mass
eigenstates and the Feynman diagram of Fig. 1 should be modified to
include this point.
In Figure 2, we explicitly showed the correction;
the incoming and the outgoing neutrinos are the mass eigenstate neutrinos and
the vertices are modified to include the mixing effects
($U_{i\alpha}$ and $U_{\alpha j}^\ast$ when the intermediate charged lepton is
$\alpha \ (e,\mu,\tau)$).
Therefore in this case one has to consider a $2\times2$ Hamiltonian matrix,
which in the mass eigenstates basis, is
\begin{equation}
{\cal H}_{ij}=U_{j\alpha}U_{i\alpha}^\ast M_{\alpha}+\frac{m_i^2}{2E}\
\delta_{ij}\ ,\
i,j=1,2,3\ ,
\label{E:3_22}\end{equation}
where $m_i$ are the running masses of the neutrino mass eigenstates.
In Eq. (\ref{E:3_22}) we are summing over $\alpha$ and $M_\alpha$ is
the value of $M$ obtained by
solving Eq. (\ref{E:3_17}) with $m$ replaced by $m_\alpha$,
the running mass of the charged lepton $\alpha$ (Eq. (\ref{E:2_1})).

In this case we find that the mixing angles change from their vacuum values.
For an illustrative purpose, we consider the case of two generations.
The mixing matrix of the neutrinos in vacuum is
	\begin{equation}
		U=\left( \begin{array}{cc} \cos\theta & \sin\theta \\
					   -\sin\theta & \cos\theta
			\end{array}
		\right)\; .
	\label{E:3_23}
	\end{equation}
The mixing angle in the medium is given in terms of the vacuum angle
$\theta$ as
	\begin{equation}
		\tan(2\theta_{_T})=
		\frac{\sin(2\theta)}
		     {\cos(2\theta)+
		{\displaystyle\Delta m^2/E\over \displaystyle(M_\mu-M_e)}}\; ,
	\label{E:3_24}
	\end{equation}
where $\Delta m^2\equiv m_2^2-m_1^2$
and $M_\mu(M_e)$ is the effective mass of $\nu_\mu(\nu_e)$, not to be
confused with the mass of the muon (electron).
For light neutrinos $\theta_{_T}$ is not much different from
the vacuum angle $\theta$ since $M_\mu-M_e$ is
much bigger than $\Delta m^2/E$.
The mass eigenstate neutrinos actually correspond to
the weak eigenstate neutrinos in the early universe
unless $\Delta m^2/E$ is comparable to the effective mass differences.

\section{ Discussion}

We have calculated the generational mass splitting of neutrinos in high
temperature
(above $250$ GeV) $SU(2)_{\scriptscriptstyle{L}}\otimes U(1)$ gauge theory
to be
$\Delta_{\alpha\beta}\equiv M_\alpha^2-M_\beta^2
	={g^2\over 32\pi^2}m^2_\alpha(T^2)\Bigl\{
	\ln\left[m^2_\alpha(T^2)/T^2\pi^2\right]+
	2\gamma_{\scriptscriptstyle{E}}
	+1\Bigr\}
	-{g^2\over 32\pi^2}m^2_\beta(T^2)\Bigl\{
	\ln\left[m^2_\beta(T^2)/T^2\pi^2\right]+
	2\gamma_{\scriptscriptstyle{E}}
	+1\Bigr\}$ where we assumed the neutrinos to be
massless at tree level and $m_\alpha(T^2)$ and $m_\beta(T^2)$
are the running masses of
the corresponding charged leptons, as given by Eq. (\ref{E:2_1}).
This effect lifts the mass degeneracy of neutrinos for temperatures
below the GUTS symmetry breaking temperature. We have also discussed the case
in which neutrinos have tree level masses and large mixing. In this
case we found that weak eigenstates and mass eigenstates are the same
unless the mass difference is large.

\acknowledgements

We wish to thank Gordon Feldman and Thomas M. Gould for helpful discussions.
C.W.Kim wishes to thank the Center for Theoretical
Physics at the Seoul National University for the hospitality
extended to him during his visit.
A. Erdas wishes to thank the High Energy Theory Group of
the Johns Hopkins University for the hospitality extended to him
during his several visits.
This work was supported in part by the National Science Foundation and
the MURST (Ministero Universit\`a e Ricerca Scientifica).


\end{document}